# Anomalous spin dynamics in charge ordered, "two-electron" doped manganite $Ca_{0.9}Ce_{0.1}MnO_3$ : consequence of a spin-liquid phase?

Ajay Sharma, Subhasis Sarangi and S. V. Bhat
Department of Physics, Indian Institute of Science, Bangalore-560012, India

The 'two-electron' doped rare earth manganites $Ca_{1-x}Ce_xMnO_3$ (x = 0.1, 0.2) are probed using resistivity, ac susceptibility and electron paramagnetic resonance (EPR) measurements across their respective charge ordering (CO) temperatures $T_{CO}$ = 180 K and 250 K. The EPR 'g' factor and intensity as well as the transport and magnetic behaviors of the two compositions are qualitatively similar and are as expected for CO systems. However, the EPR linewidth, reflective of the spin dynamics, for x = 0.1, shows a strongly anomalous temperature dependence, decreasing with decreasing temperature below $T_{CO}$ in contrast with the sample with x = 0.2 and other CO systems. Keeping in view the evidence for magnetic frustration in the system, we propose that the anomalous temperature dependence of the linewidth is the signature of the occurrence of a disorder driven spin liquid phase, present along with charge ordering.

## I. INTRODUCTION

The complex phase diagrams and the fragile phase boundaries of the hole and electron doped rare earth manganites[1,2], $A_{1-y}R_yMnO_3$, where A is a divalent alkaline earth ion ($Ca^{2+}$, $Sr^{2+}$,..) and R is a trivalent rare earth ion ($La^{3+}$, $Nd^{3+}$, $Pr^{3+}$,….), continue to attract intense theoretical[3] and experimental[4] attention. [For y < 0.5, the Mn $^{4+}$ ions (electronic configuration $3d^3$) of the parent compound $AMnO_3$ are replaced by y number of Mn $^{3+}$ ions ($3d^4$ ) with one extra electron leading to 'electron' doping and for y > 0.5, the material is considered to be doped with 'holes']. One of the activities at the center stage of this effort is the study of the fascinating phenomenon of charge ordering[5,6]. Charge ordering (CO), refers to the real space ordering of the $Mn^{4+}$ and the $Mn^{3+}$ ions which occurs on cooling the paramagnetic insulating material below a certain sample- specific temperature $T_{CO,}$ the charge ordering temperature. The CO insulating state dominated by Jahn-Teller polarons and the superexchange interaction between $Mn^{3+}$ and $Mn^{4+}$ ions and the colossal magnetoresistive (CMR) ferromagnetic metallic state dominated by the Zener double exchange (DE) interaction are understood to be two competing ground states in the phase diagram of doped manganites. The CO transition is usually accompanied by certain tell tale signatures like superlattice reflections in x-ray and neutron diffraction patterns, an increase in the resistivity and often, especially when the CO state terminates in an antiferromagnetic (AFM) phase on further cooling, a peak in the susceptibility. In recent years a large number of investigations have addressed the many interesting properties of the CO phenomenon such as the 'melting' of the CO when subjected to perturbations such as magnetic field, irradiation, and the application of a current and the influence of the size i.e., bulk vs nano[7], on the stability of the CO phase and so on.

Relatively less studied in this respect are the cerium doped managnites of the form $A_{1-x}\,Ce_xMnO_3$. It is found[8] that Ce enters into these compounds in the 4+ valence state and therefore one expects these to have certain unique properties since the "x" amount of cerium doping at the A site inducts x = 2y number of $Mn^{3+}$ ions, twice of what are induced by $Ln^{3+}$. Thus the cerium-doped oxides are expected to show a rapid change in their magnetic and electronic properties as a function of x compared to the $Ln^{3+}$ doped oxides[8].

Maignan *et. al*. prepared electron doped $Ce_xCa_{1-x}MnO_3$, CCMO, by traditional solid state route[9]. Zeng et. al.,[8] studied the magnetic and electronic properties of CCMO and compared it with the conventional electron doped $La_yCa_{1-y}MnO_3$. They gave a tentative magnetic T-x phase diagram, which agrees qualitatively well with that of LCMO. Recently Caspi et. al.[10], in a comprehensive structural and magnetic study provided a detailed understanding of the structural and magnetic phase diagram of CCMO in the region $0 \leq x \leq 0.167$. One of the noteworthy features of their finding is that for a substantial composition range ( $0.1 \leq x \leq 0.167$) the material exhibits a CO phase.

Electron paramagnetic resonance, (EPR), is a powerful local probe for the study of static and dynamic magnetic correlations on a microscopic level and can help to clarify[11-14] the complex magnetic states exhibited by manganites. The technique has proven to be especially useful to understand the CO state through the characteristic behaviours of the 'g' factor, the intensity and the linewidth across the CO transition[15-17]. It is found that in CO systems, even in the high temperature charge disordered phase, the hole doped and the electron doped systems exhibit opposite shifts of their 'g' factors consistent with the sign of their spin-orbit coupling constant of the charge carriers. Interestingly, such a shift is not observed in CMR manganites, presumably the itinerant nature of their charge carriers leading to an averaging out of the contribution of the spin orbit interaction. Below $T_{CO}$ a continuous increase in the 'g' factor is observed. Further, the EPR linewidth shows a continuous increase starting from $T_{CO}$ down to the antiferromagnetic transition to which most of the CO manganites transform on further reduction in temperature. This linewidth behaviour is explained in terms of a "motional narrowing"[16] or a "variable range hopping"[17] model.

In this work, we present our results of magnetic, transport and EPR studies on $Ca_{1-x}Ce_xMnO3$ for x= 0.1, and 0.2. Both the systems have been earlier reported[8,10] to undergo CO transitions and our experiments confirm this. The EPR signatures of CO are observed for x = 0.2. However, for x = 0.1, while the 'g' factor and the intensity behave as in x = 0.2 and other CO systems, the linewidth shows a strongly anomalous decrease with decreasing temperature below Tco. We attempt to understand this in terms of the presence of spin fluctuations that average out the linewidth broadening interactions. Such fluctuations, similar to those in a spin liquid, are consequence of the magnetic frustration present in the CCMO sample.

## II. EXPERIMENTAL

The polycrystalline $Ce_xCa_{1-x}MnO_3$ (x = 0.10, 0.20) samples were prepared by solid state synthesis[9]. Stoichiometric amounts of $CaCO_3$, $CeO_2$, and $MnO_2$ were mixed and heated at different temperatures (1100 $^0$C, 1200 $^0$C and 1350 $^0$C) with intermediate grindings. Single phasic materials were obtained only at 1350 $^0$C. The pellets were finally sintered at 1400 $^0$C. The powder X-ray patterns were recorded using a Philips diffractometer with Cu K$\alpha$ radiation and scanning (0.01 step in 2$\theta$) over the angular range $10^0$ – $110^0$. The diffractrograms for the two compositions show single phases with no impurity peaks present. The two powder patterns of could be indexed in the orthorhombic system with space group Pbnm with a ~ b ~ $c/\sqrt{2}$ . No impurity is detected by EDAX (energy dissipative X-ray analysis) either, and it gives expected cationic compositions. The oxygen stoichiometry and the $Mn^{3+}/Mn^{4+}$ ratio were confirmed by iodometric titration. The ac susceptibility measurements were done using a homemade apparatus in the temperature range of 80 K-300 K at a frequency of 100 Hz. The EPR measurements were carried out on powder samples using a commercial X-band spectrometer. The spectrometer was modified by connecting the X and Y inputs of the chart recorder to a 16-bit A/D card, which in turn is connected to a PC enabling digital data acquisition. With this accessory, for the scan width typically used for our experiments, i.e., 6000 G, one could determine the magnetic field to a precision of ~ 3 G. The temperature was varied from 10K – 300K (accuracy ± 1K) using a continuous helium flow cryostat and the EPR spectra were recorded while warming the sample. A speck of DPPH was used as a field marker to find the center field of the signal accurately.

## III. RESULTS AND DISCUSSION

The EPR signals in the two samples were quite broad, similar to those observed in other CO manganites[15-17]. For the x = 0.2 sample, as the sample is cooled, the signal further broadens, again a behaviour found in other CO manganites. However, signal from the x=0.1 sample shows an anomalous temperature dependence in that it is found to narrow down with the decrease in temperature. The signals of both the samples are fitted to the Lorentzian lineshape function given by the equation[18]

$$\frac{dP}{dH} = \frac{d}{dH}\left(\frac{\Delta H}{\Delta H^2 + (H - H_o)^2} + \frac{\Delta H}{\Delta H^2 + (H + H_o)^2}\right) \qquad (1)$$

where $H_0$ is the resonance field and ΔH is the peak to peak linewidth. The use of the two terms in the equation accounting for the clockwise as well as the anticlockwise circularly polarized component of microwave radiation is necessary because of the large width of the signals. The sharp signals due to DPPH, used as a field marker, have been digitally subtracted to aid the fitting of the lineshapes. The lineshape parameters viz. the intensity, the 'g' factor and the linewidth are extracted from the fits. Fig. 1 shows the temperature dependence of the EPR intensity for the two compositions. For both the samples, the intensity initially increases with decrease in temperature, shows a broad peak at $T_{CO}$ and decreases with further decrease of temperature in the CO phase. This behaviour is observed in other CO systems as well and also qualitatively mimics the behaviour of the susceptibility of the CO materials, shown in the insets to fig. 1 for x = 0.1 and 0.2. Neutron scattering studies on $Bi_{1-x}Ca_xMnO_3$ have shown[19] that for $T > T_{CO,}$ close to $T_{CO,}$ double-exchange induced ferromagnetic spin fluctuations are present which are progressively replaced by the superexchange mediated antiferromagnetic spin fluctuations as the system goes into the CO phase. This explains why a broad peak is seen in the susceptibility, magnetization as well as EPR measurements. The antiferromagnetic (AFM) spin fluctuations increase in the CO phase and on further cooling an antiferromagnetically ordered phase appears. This is accompanied by a decrease in the intensity of the EPR signal in the CO phase before the complete disappearance of the signal at a temperature close to $T_N$. According to Caspi et al., for $x \geq 0.1$, CCMO shows phase coexistance at low temperatures. The x = 0.10 sample undergoes a CO transition at $T_{CO}$ = 170 K, and a transition to a C-type AFM phase at 152 K and a further transition to a “magnetically charge ordered” (MCO) phase at $T_{MCO}$ = 107 K. The EPR signal disappears at 100K, close to the $T_{MCO}$. Similarly for x = 0.20, $T_{CO}$ = 250K, the EPR signal becomes very weak below 140 K and completely disappears below 90K. The temperature below which the signal becomes weak matches with the onset of C-type AFM phase. As there is still some ferromagnetic ordering present along one of the axes, the signal is observed in the C-AFM phase. In the MCO phase the signal disappears completely. Thus EPR signals give an indication of the two magnetic phases. As seen the the insets to fig. 1, which show the ac susceptibility (f = 100 Hz) for the two compositions, the CO transition temperatures observed from the ac susceptibility curves are close to those inferred from the peaks in the EPR intensities.

The temperature dependence of the ‘g’ factor, given by $h\nu = g\beta H$, where $\nu$ is the microwave frequency and H is the resonant field, for the two samples is shown in fig. 2. From room temperature down to $T_{CO}$ the g value for both the compositions is less than the free electron g value ($g_e$ = 2.0023) and is essentially independent of the temperature. For non-S state ions such as $Mn^{3+}$ and $Mn^{4+}$ the ‘g’ factor is shifted from $g_e$ and is given by $g = g_e(1- \lambda/\Delta)$ where $\Delta$ is the crystal field splitting and $\lambda$ is the spin-orbit coupling constant. The sign of $\lambda$ is positive for an electron charge carrier in a less than half filled shell. Thus the ‘g’ is expected to be less than $g_e$ for the two samples of CCMO as indeed observed in the paramagnetic phase. In the CO phase the g value increases and the increase is seen to be more in x = 0.10 sample. The CO phase is also accompanied by orbital ordering (OO) which can affect the spin-orbit coupling. In addition, the lattice parameters show a rapid change in the CO phase, which could affect the strength of crystal field interaction. The temperature dependence of the 'g' factor observed in the CO phase is a net effect of these two contributions.

Fig. 3 shows the temperature dependence of EPR linewidth ΔH(T), for the two compositions. In the paramagnetic phase the linewidth for the two samples is of the order of 2000 G. This relatively large linewidth, seen in other manganites as well, is understood to originate in the Dzialoshinsky-Moriya antisymmetric exchange interaction and the crystal field anisotropy. In the CO phase, for x = 0.20, for $T < T_{CO}$, the linewidth increases with the decrease of temperature as seen in other CO manganites. This behavior viz., the continuous decrease of the linewidth with the increase in the temperature from $T_N$ to $T_{CO}$ in the CO phase has been explained by invoking the model of motional narrowing or in terms of variable range hopping. In both of these cases, the decrease in the linewidth for increasing T from $T_N$ to $T_{CO}$ was accompanied by a decrease in the resistivity as well. In contrast, for x = 0.10, the linewidth increases with an increase in temperature in the CO phase, (*accompanied by a decrease in the resistivity*), which, to the best of our knowledge is not observed in the CO phase of any other manganite.

Thus we have this quite surprising result in the case of x = 0.1 compound that the 'g' factor and the intensity of the EPR signals as well as magnetic, transport and structural studies give evidence of a CO phase while the linewidth behaviour is exceptional. In the following we try to understand this result in terms of the present knowledge of the linewidth behaviour in manganites and other materials undergoing magnetic transitions. We find that none of these models is applicable to the present example and a new explanation is needed.

A number of reports have addressed the behavior of EPR linewidth in manganites. In manganites undergoing a transition from a paramagnetic to ferromagnetic state, the linewidth $\Delta H(T)$ in the paramagnetic state is found to increase quasilinearly with increasing T. Different models have been proposed to explain this result such as contribution from spin-phonon interaction[11], a relaxation bottleneck behavior[20,21] or in terms of a combined effect of DM and CF interactions[12]. The latter model seems to be able to explain most of the CMR EPR linewidth results and can be described in terms of the equation, $\Delta H(T) = [\chi_0(T) / \chi(T)]\, \Delta H(\infty)$ where $\chi_0(T) \propto T^{-1}$ is the free ion (Curie) susceptibility, $\chi(T)$ is the measured susceptibility and $\Delta H(\infty)$ is a temperature independent constant attributable to the high temperature limit of the linewidth. It is found that a large number of CMR manganites follow a 'universal' relation[14] when $\Delta H(T)/ \Delta H(\infty)$ is plotted against $T/T_C$. We examine the applicability of this model to the present CO systems in figure 4. According to this model, $\Delta H(T)$ should be proportional to $\chi_0 / \chi(T)$ i.e. to $1/[T \times \chi(T)$. However as clearly seen in the figure, the proportionality does not hold for either of the two CO compounds. This is really not surprising if we note that in these systems the narrowing is a consequence of the enhancement of the susceptibility as one approaches Tc. In the CCMO samples, as well as in other CO systems, for $T < T_{CO}$ , the susceptibility actually decreases thus violating the basic premise of the model.

There has been a considerable number of EPR studies[23-26] over the years on antiferrromagnetic materials above their $T_N$. In these systems, typically, as one approaches $T_N$ from above, the EPR linewidth is found to decrease either linearly or quasilinearly down to a temperature close to the transition, go through a minimum and then diverge as $T \sim T_N$. A relation of the type $\Delta H = A + BT + C\,(T - T_N)^{-p}$, which describes the observed quasilinear dependence far above $T_N$ and the critical behavior close to the transition has been often used[26] to fit ΔH(T). However, we find that ΔH(T) in CCMO as well as in other CO systems shows neither divergence close to $T_N$, nor a quasilinear behavior far away from $T_N$. Therefore, it is unlikely that the models used to explain the ΔH(T) behavior in other AF materials is are applicable to CO systems.

In most CO materials in the CO state, ΔH(T) decreases with increasing T and in as much as it is associated with a decrease in the resistivity, hopping of the spins along with the charge carriers (i.e. holes or electrons) was understood[16,17] to cause averaging out of the broadening interactions such as DM and CF, leading to a narrowing of the linewidth. However, while the ΔH(T) for x = 0.2 may be explained according to one of these models, the ΔH(T) for x = 0.1 obviously cannot be explained in terms of either of these models because for this compound the results on ΔH(T) and ρ(T) where ρ is the resistivity, with decreasing temperature below $T_{CO}$, while charge dynamics slows down, spin dynamics becomes faster.

According to the comprehensive study of CCMO by Caspi et al.,[10] for x ~ 0.1, with decreasing T, the sample undergoes a transition from a monoclinic, paramagnetic phase to a monoclinic orbitally ordered C-type phase. As mentioned before, their resistivity, X-ray structure and ac susceptibility measurements indicate that this phase is also charge ordered. This conclusion is further supported by our resistivity, ac susceptibility and EPR ‘g’ factor and intensity behaviours. For x ≥ 0.125, all indicators mentioned above, both in the studies of ref. [10] and ours provide the evidence for a transition from room temperature paramagnetic monoclinic phase to a low temperature charge ordered phase. Indeed Caspi et al.,[10] conclude that the CO phase is of Wigner crystalline type as well. However, quite interestingly, they are not able to see CO peaks in their neutron diffraction study though they observe CO signatures in their X-ray experiments. More importantly, they provide evidence for frustration in part of the Mn-O bonds in the CO structure. The concentration of the frustrated bonds is found to increase with decreasing x. Thus both disorder and frustration being present, it is quite likely that the spin system acquires the possibility of entering into either a spin glass or a spin liquid state[27]. Considering the relatively high temperature range of the experiment, the latter appears to be the more likely scenario. This proposed spin liquid state is also qualitatively consistent with the recent work of Huber[28] who finds that for a geometrically frustrated Heisenberg antiferromagnetic system, for T approaching $T_N$ from above, below the susceptibility peak there is a rapid increase in the spin diffusion constant with decreasing temperature as well as a decrease in the spin-spin correlation time. Both these effects would go towards decreasing the EPR linewidth. We note, however, that in the present case one is concerned with the peak in the susceptibility at $T_{CO}$. Moreover, the system is also quite anisotropic, though the spin system if it were in the liquid phase would impart the isotropy inherent in Huber's treatment. Clearly more theoretical work is required along these lines as well as experimental investigations using other techniques such as inelastic neutron scattering, especially looking for the presence of any diffuse scattering.

Summarizing, we have studied the 'two-electron' doped rare earth manganites $Ca_{1-x}Ce_xMnO_3$ (x = 0.1, 0.2) using resistivity, ac susceptibility and electron paramagnetic resonance (EPR) measurements across their charge ordering (CO) transition temperatures. While for x = 0.2, all measurements are consistent with the CO phase, for x = 0.1, the EPR linewidth alone, which is an indicator of spin dynamics, shows anomalous behaviour. We examine this result in terms of the existing theories and models of EPR linewidth in manganites and other materials undergoing magnetic transitions and find that they are not able to explain this anomalous linewidth behavior. Further, the more recent qualitative models proposed for CO manganites also are found to be not applicable. Therefore we put forward a proposal, which needs to be confirmed by other techniques, that the anomalous temperature dependence of EPR linewidth in $Ca_{0.9}Ce_{0.1}MnO_3$ could be due to the presence of a spin liquid state.

## FIGURE CAPTIONS

**Figure 1:** EPR intensity plotted as a function of temperature for the two compositions. The insets show the ac susceptibility. Arrows indicate the charge ordering transition temperatures.

**Figure 2:** The ‘g’ factor plotted as a function of temperature for the two compositions.

**Figure 3:** The epr linewidth plotted as a function of temperature for the two compositions. The insets show the temperature dependences of the corresponding resistivities.

**Figure 4:** Test for the applicability of the model used for fitting the EPR linewidths of CMR materials to the present data. Circles correspond to x = 0.1 and stars to x =0.2. The dashed and the solid lines are linewidths divided by a scaling factor (right ‘y’ axis),

for the 0.1 and 0.2 sample respectively. Inset shows the relevant region on an expanded scale. The inapplicability of the model in the CO region is clearly seen.

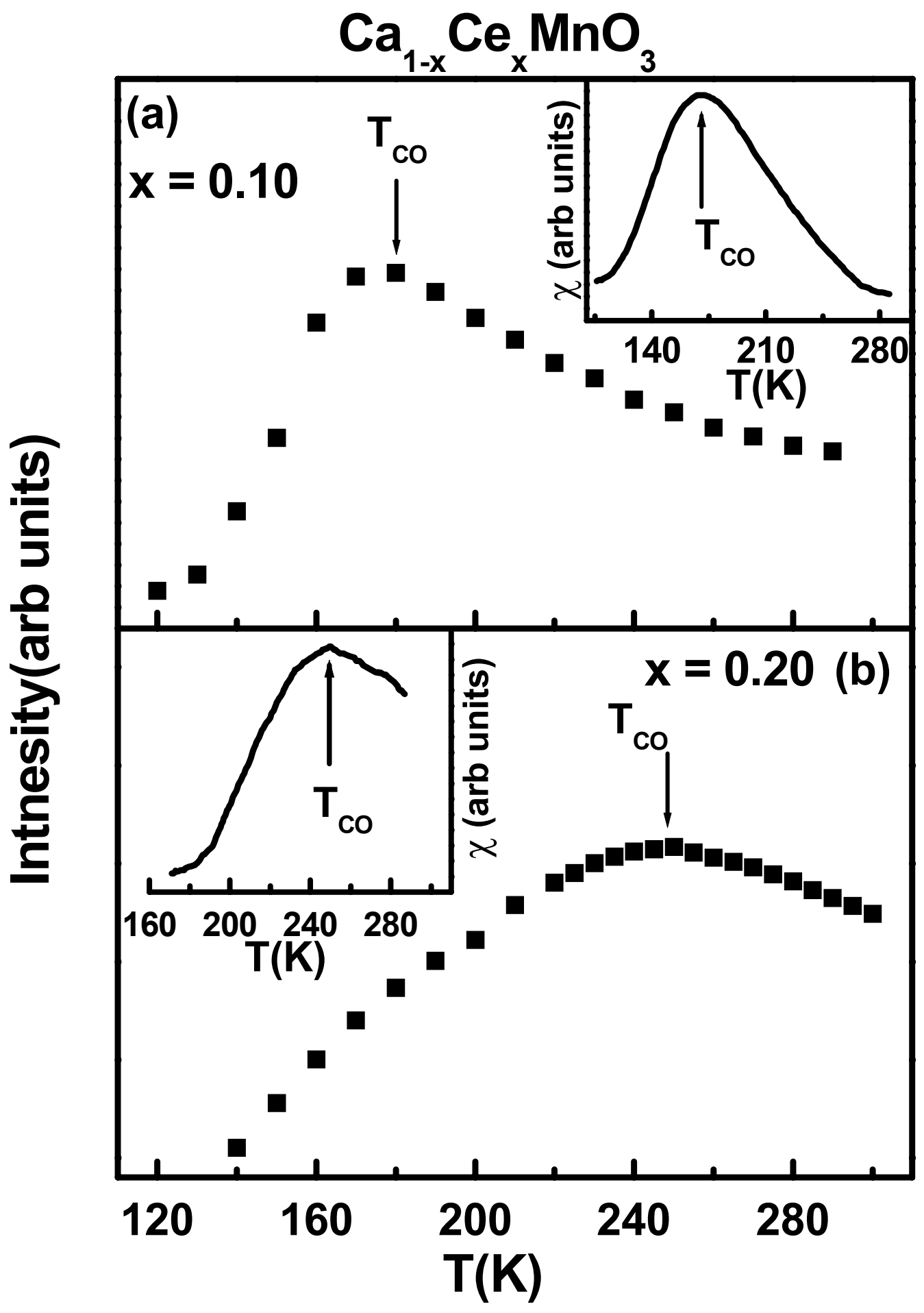

**Figure 1: Sharma, Sarangi and Bhat**

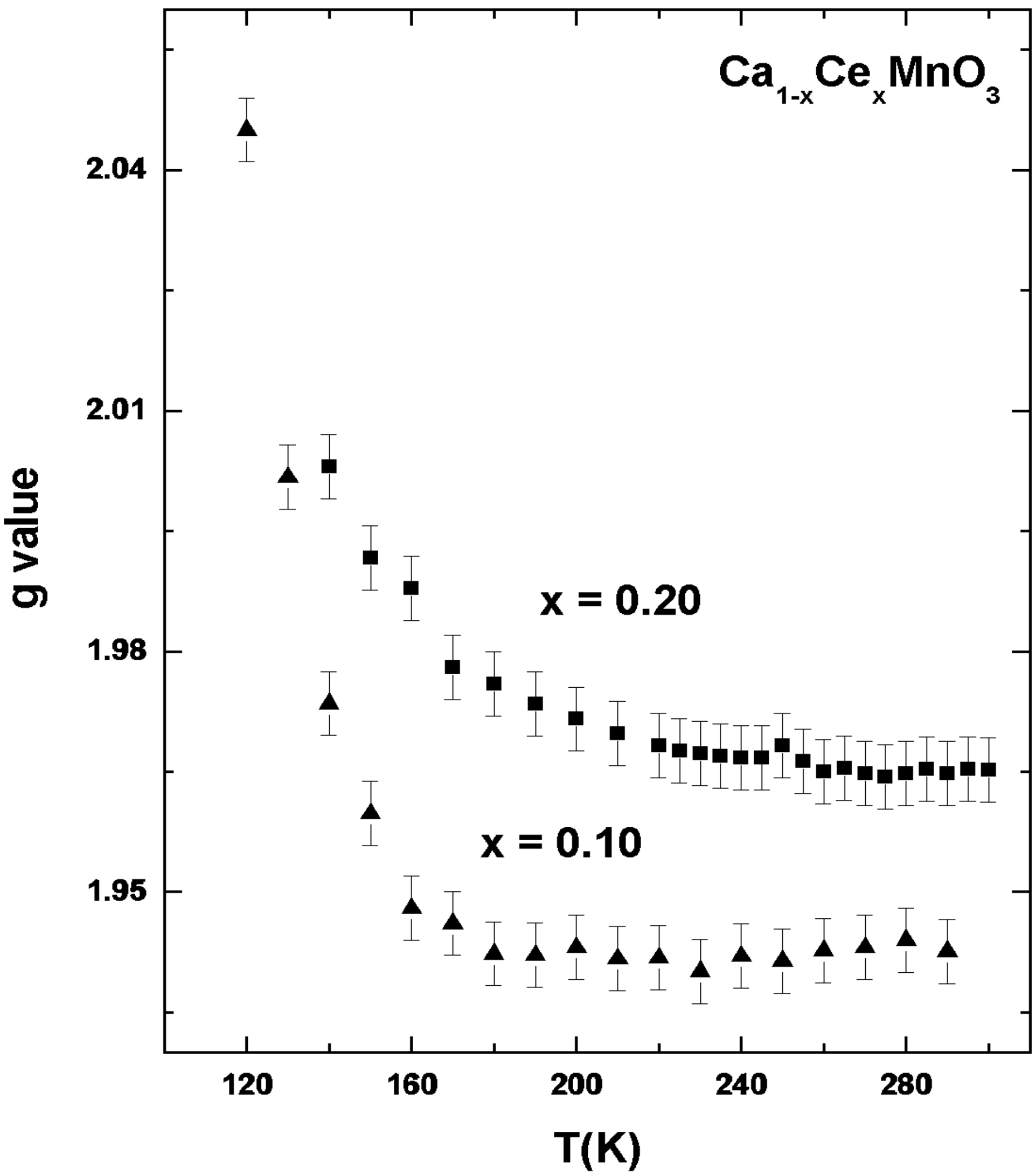

**Figure 2: Sharma, Sarangi and Bhat**

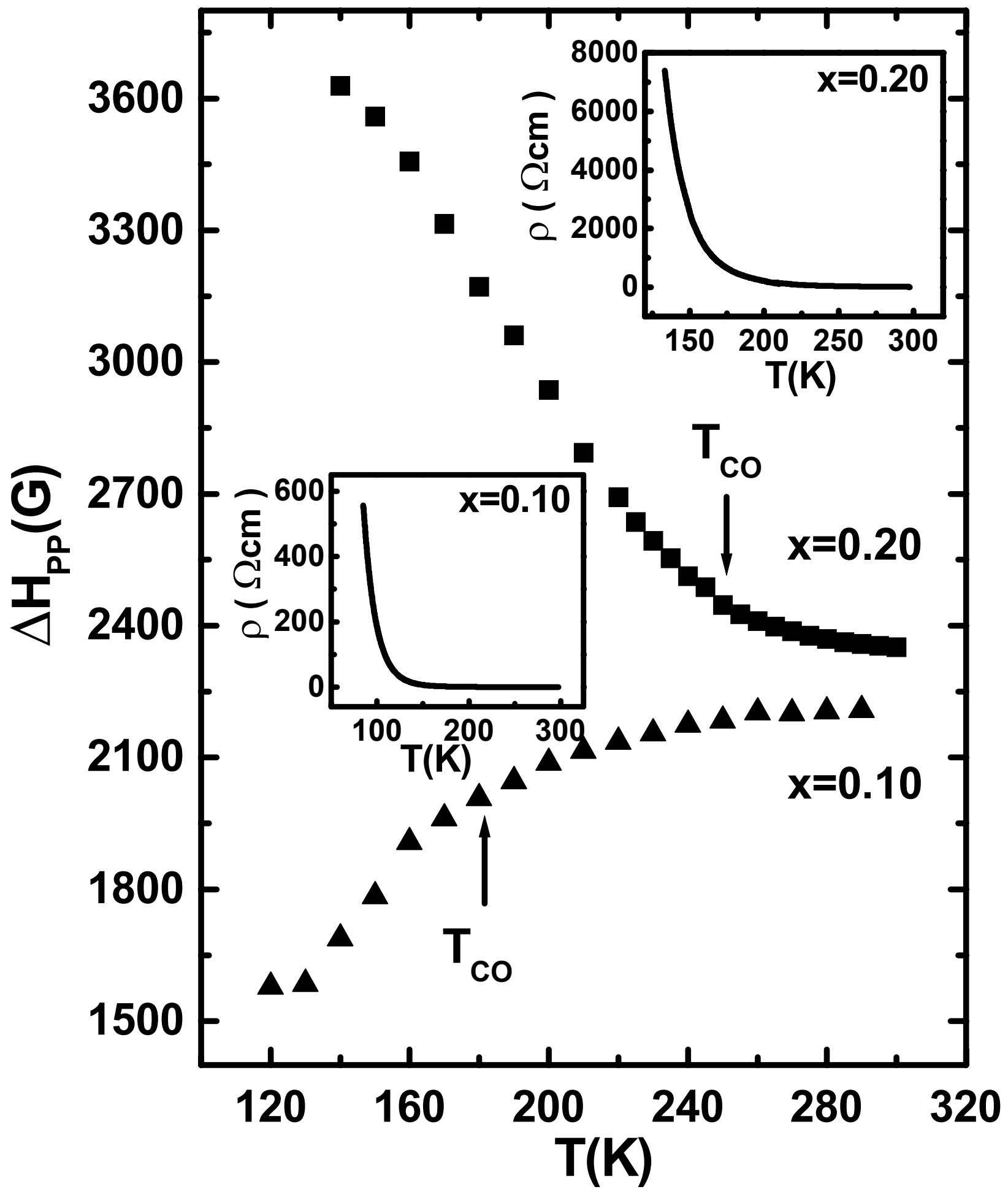

**Figure 3: Sharma, Sarangi and Bhat**

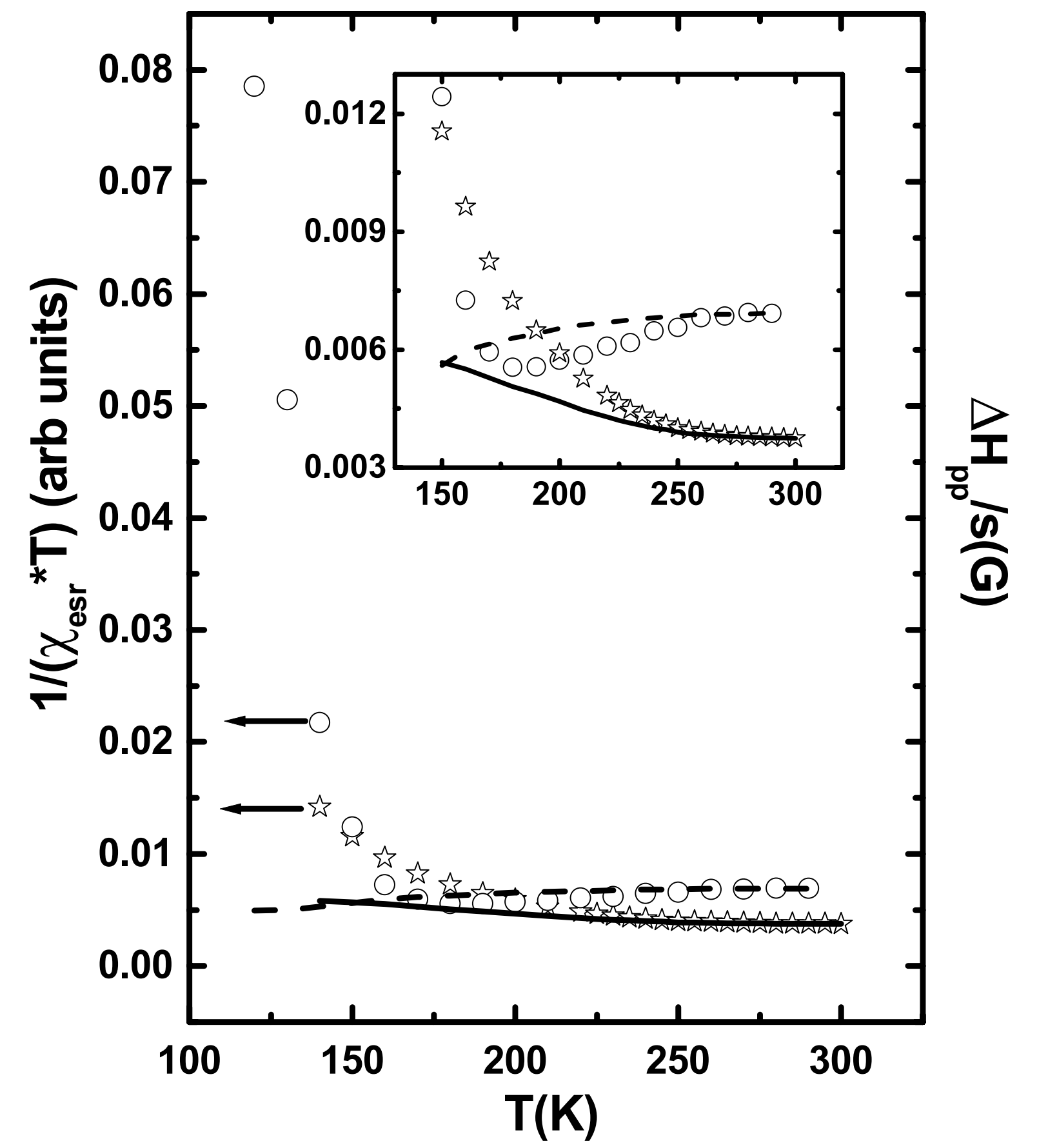

**Figure 4: Sharma, Sarangi and Bhat**